\documentstyle[prd,aps,epsfig]{revtex}

\def\be{\begin{equation}}
\def\ee{\end{equation}}

\begin{document}  
\draft
\twocolumn[\hsize\textwidth\columnwidth\hsize\csname
@twocolumnfalse\endcsname
\preprint{ }


\title{P, C and Strong CP in Left-Right Supersymmetric Models}

\author{Rabindra N. Mohapatra$^a$, Andrija Ra\v{s}in$^b$ and 
Goran Senjanovi\'c$^b$}

\address{{\it $^a$Department of Physics, University of Maryland,
College Park, MD 21218, USA  \\
$^b$International Center for Theoretical
Physics, 34100 Trieste, Italy }}

\maketitle

\begin{abstract}

We systematically study the connection between P, C and strong CP
in the context of both non-supersymmetric and supersymmetric 
left-right theories. 
We find that the solution to the strong CP problem
requires both supersymmetry and parity breaking scales
to be around the weak scale. 

\end{abstract}
\pacs{hep-ph/9707281 \hskip 1 cm UMD-PP-97-118}
\vskip1pc]

{\bf A. Introduction}\hspace{0.5cm} 
There are two possible ways of solving the strong CP problem. The first
is the dynamical relaxation mechanism, such as the celebrated
Peccei-Quinn symmetry which promotes the strong CP phase into a 
dynamical variable\cite{peccei}.

The second idea is to utilize some discrete
symmetry\cite{all,segre,nelson} to make
the strong CP phase vanish at the tree level. It then becomes calculable in 
perturbation theory and a viable solution to the problem requires that
these perturbative corrections be below the experimental upper limit.
The most appealing candidates for this job are
the fundamental spacetime symmetries: parity (P) and time reversal (CP).

Rather natural candidates are left-right (LR) symmetric theories\cite{ps74} 
which provide a framework for the spontaneous breakdown of parity. Furthermore,
CP can be spontaneously broken even in the minimal version of
these theories\cite{gs}. In addition they can be embedded in SO(10) grand
unified theories, which are the minimal truly unified models of
quarks and leptons. In this letter, we focus our attention on these
natural candidates for the solution of the strong CP problem, both in
ordinary and supersymmetric versions\cite{mr96}.

It is well known that the strong CP problem contains two aspects, 
that is the smallness of $\theta_{\rm QCD}$, 
the coefficient of the $F \tilde{F}$ term,
and the smallness of $\theta_{\rm QFD} = {\rm ArgDet} {\bf M}$, 
where 
${\bf M}$ is the mass matrix of colored fermions. It is highly suggestive to 
use parity since it implies both 
$\theta_{\rm QCD} = 0$ and ${\bf M} = {\bf M}^\dagger$ which in turn gives
$\theta_{\rm QFD} = 0$. 
 
This would be sufficient if parity were an exact symmetry of nature.
However, parity must be broken and the real challenge in these theories
is to keep $\theta \equiv \theta_{\rm QCD} + \theta_{\rm QFD}$
small to all orders in perturbation theory. Without supersymmetry this is 
an impossible task. Essentially, the problem is that the requirement of
weak CP violation destroys the hermiticity of the quark mass matrices
already at the tree level which induces in general large  
$\theta_{\rm QFD} = 0$. Another way to see it is to note that the constraint
of parity invariance alone allows for complex couplings in the Higgs potential
which lead to complex VEVs for the Higgs fields and thereby destroy the
hermiticity of the quark mass matrices even at the tree level. 
Recently it has been argued\cite{mr96} that making the left-right symmetric
model supersymmetric leads to a Higgs potential where all coupling
parameters are real thus giving us a CP-conserving vacuum.
Furthermore, the perturbative one-loop
contributions to $\bar{\theta}$ can be shown to be small
under certain circumstances\cite{mrdrugi,posp}.

 These observations have inspired us to revisit the SUSYLR model
and carefully discuss under what circumstances
$\bar{\theta}$ in this model
is guaranteed to be acceptably small. 
We find that supersymmetry and parity symmetry by
themselves are not sufficient to control the one loop contributions.
One needs charge conjugation invariance (C) for the purpose.
It then turns out that in general the hermiticity of the quark mass matrices
can only be preserved at the expense of weak CP 
violation thus making the theory unrealistic. 
We find one exception: {\it low scale of parity breaking $M_R$
and parity breaking achieved only through nonrenormalizable operators}. In this
case the smallness of $\bar{\theta}$ is achieved by a soft violation
of CP and is controlled by the small ratio of $M_R / M_{Planck}$. 
We find it rather interesting that the requirement of smallness of the
strong CP phase requires experimentally accessible scale of 
parity restoration. This is the major new result of our paper.

In order to set the framework for our discussion we first analyze 
the essential features of parity and charge conjugation and their role
in the strong CP problem.


{\bf B. No supersymmetry} \hspace{0.5cm}
We start with the minimal left-right symmetric theory based on the
gauge group $SU(2)_L \times SU(2)_R \times U(1)_{B-L} \times P$ and the 
following fermionic content: 
\begin{eqnarray}
Q_L & = & \left( \begin{array}{c}                
u \\ d \end{array} \right)_L \, (2,1,1/3) \, , \,
Q_R = \left( \begin{array}{c} 
u \\ d \end{array} \right)_R \, (1,2,1/3) \, , \nonumber\\
L_L & = & \left( \begin{array}{c}                
\nu \\ e \end{array} \right)_L \, (2,1,- 1) \, , \,
L_R = \left( \begin{array}{c}                
\nu \\ e \end{array} \right)_R  \, (1,2,- 1) \, , \,
\end{eqnarray}
with gauge quantum numbers spelled out in brackets.


Under parity these fields transform as usual 
\be
Q_L \leftrightarrow Q_R \, ; \, L_L \leftrightarrow L_R 
\ee
and similarly under charge conjugation
\be
Q_L \leftrightarrow (Q^c)_L \equiv C {\bar{Q}}^T_R \, ; \,
L_L \leftrightarrow (L^c)_L \equiv C {\bar{L}}^T_R \, .
\ee

We will stick to the somewhat conservative assumption that there are no new
quarks and leptons. It
should be mentioned that if this assumption 
is relaxed it is possible to construct
viable models based on parity only which predict calculably small theta 
\cite{bm,bcs,carlson}. Similarly, with additional fermions,
one can use both P and C symmetries and CP can be used to exchange the
two SU(2) groups \cite{lavo97}. 

With the above fermion content the field that provides quark
and lepton masses is the Higgs bidoublet
\begin{equation}
\phi \, (2,2,0) \, ,
\end{equation}
which under parity transforms as
\begin{equation}
\phi \leftrightarrow \phi^\dagger \, .
\end{equation}
Keeping in mind eventual SO(10) embedding, we allow a sign ambiguity
in the charge conjugation transformation of $\phi$
\begin{equation}
\phi \leftrightarrow  \pm \phi^T \, .
\label{ctransf}
\end{equation}

The imposition of parity and the gauge symmetry determine
the Yukawa couplings
\begin{equation}
L_y = {\bf h}_q \bar{Q}_L \phi Q_R +
{\bf h}_l \bar{L}_L \phi L_R
\end{equation}
to be hermitean i.e.
\begin{equation}
{\bf h}_q = {\bf h}_{q}^\dagger \, ; \,
{\bf h}_l = {\bf h}_{l}^\dagger \, .
\end{equation}
Clearly, since the quark mass matrices are given by
\begin{equation}
{\bf M}_q = {\bf h}_q <\phi> \, ,
\end{equation} 
they will be hermitean if and only if $<\phi>$ is real ($<\phi>$
real obviously preserves parity). But $<\phi>$ can be real only if
the Higgs potential is CP conserving. 
Here lies the crux of the
problem. Now, ${\bf h}$ is either real or complex. If it is real
then $<\phi>$ must be complex in order for CP to be
broken, in which case ${\bf M}_q$ cannot be hermitean. 
If on the other hand, ${\bf h}$ is complex
then unfortunately there are complex couplings in the Higgs potential
and $<\phi>$ itself becomes complex, destroying again the 
hermiticity of ${\bf M}_q$. 

Let us demonstrate this in some detail. Consider first the minimal case
with a single bidoublet $\phi$. It is a simple excercise to show that the
potential which depends on $\phi$ only has all the couplings real due to
parity symmetry. However, 
in order to break $SU(2)_R$ symmetry at the scale 
$M_R >> M_W$ we need other Higgs fields, $\chi_L$ and $\chi_R$,
which are nontrivial representations under $SU(2)_L$ and $SU(2)_R$,
respectively. The troublesome couplings in the schematic representation
are
\be
(\alpha \chi_L^\dagger \chi_L 
+ \beta \chi_R^\dagger \chi_R)  \phi^\dagger \phi + h.c. \, .
\ee
Parity imposes only $\alpha = \beta^*$ so that $\alpha$ is in general
complex. Only if one imposes {\it charge conjugation on top of parity}, are
these couplings made real. As we will see, this additional requirement
of C-invariance in addition to parity happens also in the supersymmetric
version. Now however, $<\phi>$ must be complex
in order to have nonvanishing weak CP violation since C-invariance
also makes the Yukawa couplings real. The hermiticity  
of ${\bf M}_q$ required for $\bar{\theta}$ to vanish is then lost.

One could imagine a possible way out along the lines of Ref.
\cite{mrdrugi}.
Suppose that there are two bidoublets with opposite transformation
properties under C 
\begin{equation}
\phi_1 \rightarrow \phi_1^T \, , \,
\phi_2 \rightarrow -\phi_2^T \, .
\end{equation}
This implies that 
$h_1 = h_1^T$ and real, and $h_2 = -h_2^T$ and purely imaginary.
In the context of SO(10), $\phi_1$ would belong to 10-dimensional
representation and $\phi_2$ to 120-dimensional one. It is noteworthy
that in SUSY one {\it must} have at least two bidoublets in order to have
nonzero quark mixing angles.

Now the quark mass matrices become
\be
M_q = h_1 <\phi_1> + h_2 <\phi_2> \, .
\ee
Notice that $<\phi_2> \rightarrow - <\phi_2>^*$ under CP, so that 
real $<\phi_2>$ breaks CP invariance.
Obviously if both $<\phi>_i$ are real, M is hermitean and complex.
This would guarantee weak CP violation without the strong one.
At this point all seems well since as before
the interaction terms between the $\phi$s in the potential
are real. However, again there are complex couplings with 
$\chi_L$ and $\chi_R$ fields of the type
are
\be
i (\alpha \chi_L^\dagger \chi_L 
+ \beta \chi_R^\dagger \chi_R ) \phi_1 \phi_2 + h.c. \, .
\ee
Parity imposes $\beta = - \alpha^*$, and charge conjugation makes $\alpha$
real. Obviously, the presence of both real and imaginary
couplings in the potential will render the VEVs of the bidoublets complex.
This in turn kills the hermiticity of the mass matrices and implies a
strong CP phase already at the tree level. We should stress that this
problem is generic and does not depend on the choice of $\chi$ fields,
{\it i.e.} whether they are doublets, triplets or higher representations.

In supersymmetry it is the superpotential that defines the theory and
one might hope that at least at the renormalizable level such dangerous
terms may be absent \cite{mr96}. However, the issue is more subtle and now
we discuss it in detail.



{\bf C. Supersymmetry} \hspace{0.5cm}
It is well known that in supersymmetry one needs at least two bidoublets
to get realistic fermion mass matrices 
so that the above scenario finds here its natural place.
There are however new
CP problems in supersymmetry: the relevant one for us is that
the masses of gauginos are complex in general. 
Here P and C play again a fundamental role: P makes
gluino mass real but not the masses of the left and right winos.
At the one loop level, these complex masses lead to a finite but
unacceptable contribution to $\bar{\theta}$ of order $\alpha/4 \pi$.
In Ref.\cite{mrdrugi},  one appeals to SO(10) grand unified extension
in order to make these masses real. The point is simply that parity
and charge conjugation suffice: P makes gluino mass real, and P and 
C ensure the same for weak gaugino masses. Thus we impose both
of them and study the consequences.

Interestingly enough, we find that
complete consistency of the theory requires that the $W_R$ mass must be in the
TeV range.

Let us go back to Eq. (11). 
In the minimal left-right model with the see-saw\cite{seesaw}
mechanism the
$\chi$ fields are taken to be triplets $\Delta$ and $\Delta_c$
under $SU(2)_L$ and $SU(2)_R$ groups respectively\cite{ms80}. Of
course
anomaly cancellation in supersymmetry requires the doubling of such fields
($\overline{\Delta}$ and $\overline{\Delta}_c$). 
It is easy to see that at the renormalizable level there are no such dangerous
complex couplings in the superpotential. The problem is that in this
model there can be no spontaneous breakdown of left-right symmetry
and if the theory is augmented by the parity odd gauge singlet field,
this gets cured at the expense of the breakdown of electromagnetic
charge invariance\cite{km94}. The way out of this impasse is to either include
more Higgs fields \cite{km94,abs} or to resort to nonrenormalizable dimension
four terms in the superpotential\cite{mr96}. 
In either case one necessarily generates
imaginary couplings described above.

Now we discuss these cases step by step starting with the renormalizable
superpotential.

(a) If one adds B-L neutral
triplets $\Omega$ and $\Omega_c$ one finds a consistent and realistic
theory with a possibility of phenomenologically interesting low
B-L scale without the need for the parity-odd singlet\cite{abs}. 
However, in this theory the new terms in the 
superpotential (we are only schematic here in notation; for exact 
expressions see \cite{abs})
\be
i \, \alpha \Omega \phi_1 \phi_2 + i \, \beta \Omega_c \phi_1 \phi_2
\label{12}
\ee
are of the type discussed above and thus $\alpha =
- \beta$ real. Next, it can be shown  
that $\Omega_c$ VEV is real. The terms in the superpotential
which are relevant are the couplings of the right handed triplet 
fields
\begin{eqnarray}
W & = &
m_{\Delta} ( Tr(\Delta \overline{\Delta}) +
Tr(\Delta_c \overline{\Delta}_c ) ) \nonumber\\
 & + &  a ( Tr (\Delta \Omega \overline{\Delta} )
+ Tr (\Delta_c \Omega \overline{\Delta}_c ) ) \, .
\end{eqnarray}
It can be easily seen that C and P render the above couplings
real. In the P breaking and electromagnetic charge preserving vacuum
$<\Delta>=<\overline\Delta>=<\Omega>=0$ and
$<\Omega_c>= M_R diag(1,-1)$ with $M_R$ being a real number
\begin{equation}
M_R = { m_\Delta \over a} \, .
\end{equation}
This induces the imaginary $\mu$ type effective mixing term between
$\phi_1$ and $\phi_2$, thus making it impossible to keep both
bidoublet VEVs real. This just as in the nonsupersymmetric case
destroys the hermiticity of the quark mass matrices. 

(b) Alternatively, one can work without $\Omega$ fields assuming
that there are nonrenormalizable terms in the superpotential
to achieve the spontaneous breakdown of parity. Again, one can
do without the parity-odd singlet \cite{ams}. In this case
the analog of the mixing of $\Omega$ and $\phi$ fields 
(\ref{12}) is achieved through following d=4 terms
in the superpotential
\be
i \, {\alpha \over M_{Pl}} \Delta \overline{\Delta} \phi_1 \phi_2 
+ 
i \, {\beta \over M_{Pl}} \Delta_c \overline{\Delta_c} \phi_1 \phi_2 \, .
\ee
Again $\alpha = -\beta$ is real. Now clearly the complex mixing term
between $\phi_1$ and $\phi_2$ is suppressed by $M_R \over M_{Pl}$.
It is easy to see that the relative phase between the 
$\phi_1$ and $\phi_2$ VEVs can be controlled by the same 
suppression factor. It is a simple excercise to show
that the strong CP phase is of order
\be
\theta = { {M_R^2} \over {m_S M_{Pl}} } \, ,
\ee
where $m_S$ is the scale of SUSY breaking in the light
particle sector of the theory. At this level, clearly
this parameter is completely undetermined. 

On the other hand in this version of the theory, the 
splitting of the bidoublets is achieved through the
above d=4 terms and thus besides the usual two light
doublets of the MSSM above the scale $M_R^2 / M_{Pl}$ there
will appear the other two doublets. It has been shown
in Ref. \cite{mrdrugi} that the running of the Yukawa couplings 
below $M_R$ quickly generates sizeable $\theta$ 
when four doublets are present. 

Thus one is forced to the low parity breaking scale scenario. One way to
get this is to introduce a parity odd singlet 
superfield $\sigma$ \cite{km94}. In this case, in order not to introduce
complex couplings in the superpotential an additional parity even 
singlet $X$ is needed. Namely, a parity odd singlet has
imaginary couplings with the bidoublets and thus one must insure
that its VEV be imaginary too. This can be achieved if one chooses a
superpotential for the singlets of the form 
\be
W_s = X ( \alpha \sigma^2 + M^2 ) \, ,
\ee
where $\alpha$ and $M^2$ are real by parity.

Those who dislike model building
should know that it is possible to do away with singlets completely.
Instead, one can obtain a desired
pattern of symmetry breaking using only nonrenormalizable operators,
as long as the neutrino Yukawa couplings satisfy
$ f \leq 10^{-2} - 10^{-3}$ and $ m_S \approx M_R \approx 1$ TeV.
Namely, in this case the nonrenormalizable terms should lower the 
energy at the parity broken extremum to be the minimum of the
potential. The couplings $f$, due to running between the
scale $M_U$ of assumed universality of soft terms and the scale
of right handed neutrinos $M_{\nu_R}$, cause the difference
between the VEVs 
$v = <\Delta_c> $ and $\overline{v} = <\overline{\Delta}_c>$
\be
v^2 - \overline{v}^2 \approx { f^2 \over { 16 \pi^2} }
ln ( {M_U \over M_{\nu_R} }) m_S^2 \, .
\ee
We find that the condition for the parity breaking and electric charge
conserving minimum is
\be
{g^2 \over 2} (v^2 - \overline{v}^2 )^2 <
{m_{\Delta} \over M_{Pl}} v^2 \overline{v}^2
\ee   
for a range of values of parameters that characterize the non-renormalizable
terms in the superpotential. The condition on $f$ noted above follows from this 
inequality.

Next, in order to break CP we need nonzero (and real) VEVs of both
bidoublets. This can only happen if there is a mixing term between
$\phi_1$ and $\phi_2$. This term (real due to parity) breaks C softly. 
This result is a reflection of a general theorem regarding the
impossibility of spontaneous CP violation in the SSM with four Higgs
doublets\cite{masi95}. If one does want to stick to spontaneous
violation, this is easily achieved with two singlets as in the above
example. 
In the presence of this soft C-breaking term, one expects finite
contributions to the phase of the left and right gaugino masses. There are
no one loop contributions to such phases. If they arise at the two or
higher loop level, their contribution to the $\Theta$ is $\leq \alpha^3/
(4\pi)^3$ which is of order $10^{-9}$ and is therefore small. Also
we repeat that there is a one loop contribution to quark masses due to
the soft SUSY breaking terms that has already been evaluated in Ref. 9
and is shown that it can be around $10^{-9}$ to $10^{-10}$ level.


{\bf D. Summary and outlook} \hspace{0.5cm}
The main implication of our work is the low scale of parity breaking,
necessary for the solution to the strong CP problem. Let us
briefly comment on the implications of this model for neutrino masses.
The smallness of the neutrino masses in our model is of course
guaranteed by the see-saw mechanism. As far as the values of the neutrino
masses are concerned, it depends on the precise model for the Dirac
neutrino masses in the theory. In order for the neutrino masses to be 
below the upper experimental bounds, one must assume neutrino Dirac mass
terms order of magnitude or so smaller than the charged lepton
masses.  This in turn implies that 
$\nu_{\mu}$ and $\nu_{\tau}$ have to decay and both the $\nu_{\tau}$ and
$\nu_{\mu}$ can decay only through the exchange of the neutral 
component of the left-handed triplet $\Delta$ \cite{rs81} rapidly enough
to satisfy necessary cosmological constraints. 
Although in nonsupersymmetric version of the theory this requires
some mild fine tuning since the mass of the left handed triplet
is proportional to $M_R$\cite{hn87}, in the supersymmetric version
discussed here this does not happen. This scenario is phenomenologically
completely consistent\cite{mnz94} and has interesting predictions of rare $\mu$
decays and $\mu-\overline{\mu}$ conversions\cite{mp86}.
Another possibility for being in accord with cosmological limits on
neutrino masses is to suppress the neutrino Dirac mass terms as a higher
order loop effect\cite{bm91}. The model in this case has to be supplemented
by the addition of extra color triplet fields coupling to quark fields, which
do not affect the discussion of the strong CP problem given above.

It is well known that in left right models with low $M_R$, there are
tree level neutral Higgs contributions to the flavor changing neutral
current effects. Present observations require that the mass of these
neutral Higgs bosons must be more than 5 TeV or so. Since these masses
are proportional to $M_R$, this is consistent with our result that
that puts $M_R$ also in the same TeV range. 

Since our results heavily depend on the imposition of charge conjugation
on top of parity it is natural to consider the SO(10) GUT extension of
LR models. Namely, in SO(10) charge conjugation is an automatic 
gauge symmetry and furthermore, as we remarked before, our choice
of the C-transformation properties of bidoublets would simply imply
that $\phi_1$ lies in the 10-dimensional, and $\phi_2$ lies in the 
120-dimensional representation. On the other hand, it is hard, if not 
impossible, to achieve low $M_R$ in the supersymmetric SO(10), at least 
in the minimal version of the theory.

In conclusion, we stress that this is a natural solution to the 
strong CP problem since low $M_R$ scale (order $m_S$) can be achieved 
naturally in the process of minimization of the
potential. Consistency with the hierarchy problem suggests
then $M_R$ of order few TeV's. We find it rather appealing
that the smallness of $\theta$ in left-right symmetric 
theories is linked to both supersymmetry and $M_R$ being
at the low scale. This provides to date the strongest theoretical 
motivation for a low mass
$W_R$ which has long been of great phenomenological and experimental
interest.

The work of R. N. M. is supported by the National Science Foundation
grant no. PHY-9421386 and the work of A. R. and G. S. partially by
EEC grant under the TMR contract ERBFMRX-CT960090.

\end{document}